\documentclass{aa}
\usepackage{graphicx}
\usepackage{natbib}
\usepackage{txfonts}
\newcommand{\HI}{\ion{H}{i}}
\newcommand{\HII}{\ion{H}{ii}}
\newcommand{\kms}{km sec$^{-1}$}

\newcommand{\cmtwee}{cm$^{-2}$}

\begin{document}
   \title{A large \HI\ cloud near the centre of the Virgo cluster}
\authorrunning{Oosterloo \& van Gorkom}

   \author{Tom Oosterloo
          \inst{1}
          \and
          Jacqueline van Gorkom\inst{2,3}         
}


   \institute{Netherlands Foundation for Research in Astronomy, 
              Postbus 2, 7990 AA, Dwingeloo, The Netherlands\\
              \email{oosterloo@astron.nl}
         \and
             Department of Astronomy, Columbia University, 
              550 West 120th Street, New York, NY 10027
        \and
             Kapteyn Astronomical Institute, University of Groningen, 
              Postbus 800, 9700 AV Groningen, The Netherlands
 }

    \date{Received:  Accepted:}

   \abstract{ We report the discovery of a large \HI\ cloud in the central
   regions of the Virgo cluster.  It is $110\times 25$ kpc in size and contains
   $3.4\times 10^8\ M_\odot$ of \HI. The morphology and kinematics of this
   cloud strongly suggest that it consists of \HI\ removed from the galaxy NGC
   4388 by ram-pressure stripping. It is more likely the result of an
   interaction of the ISM of NGC 4388 with the hot halo of the M86 group and
   not with the ICM centred on M87. The large extent of the plume suggests
   that gas stripped from cluster galaxies can remain
   neutral for at least $10^8$ yr.  Locally, the column density is well above
   $10^{20}$ \cmtwee, suggesting that the intra-cluster \HII\ regions known to
   exist in Virgo may have formed from gas stripped from  cluster
   galaxies. The existence of the \HI\ plume suggests that stripping of infalling
   spirals contributes to the enrichment of the ICM. The \HI\ object in the
   Virgo cluster recently reported by Minchin et al.\ may have a similar
   origin and may therefore not be a ``dark galaxy''.
 }

   \maketitle
%

\section{Introduction}

It is well known that clusters are hostile environments for galaxies and that
galaxies in clusters differ in many aspects from field galaxies.  Due to the
high galaxy density, interactions between galaxies occur frequently, affecting
and transforming them.  Moreover, cluster galaxies move at high speed through
the intra-cluster medium (ICM) and this ICM can remove gas from the galaxies
by ram-pressure stripping. One very clear manifestation of these effects is
that the properties of the neutral hydrogen of cluster spirals are severely
affected by cluster related processes.  Many cluster spirals contain
(sometimes much) less neutral hydrogen in comparison to what expected for
their type and luminosity \citep[e.g.][]{cay90,sol01}.
\HI\ imaging studies of such galaxies show that their \HI\ disk is often
truncated to well inside the optical disk
\citep[e.g.][]{war88,cay90}. 

The processes responsible for gas removal (tidal interactions and ram-pressure
stripping by the ICM) have been studied in detail, both theoretically
\citep[e.g.][]{aba99,qui00,sch01,vol01} and observationally
\citep[e.g][]{war88,cay90,sol01,ken04}.  Nevertheless, questions remain.  One source
of uncertainty is that the orbit of a galaxy through the cluster is unknown so
that assumptions have to be made about where and when the stripping occurred.
Even less clear is the fate of the gas once it has been removed from a cluster
galaxy and has ended up in intra-cluster space.  Being embedded in the hot
cluster halo, one might expect the stripped gas to evaporate and enrich the
cluster halo \citep[e.g.][]{vei99,vol01}.  However, the timescale for this
evaporation is uncertain, e.g.\ because it depends on the structure of the
magnetic field. Moreover, if the column density of the stripped gas is
sufficiently high, it may condense into clouds of mostly molecular gas with
only a skin of neutral hydrogen \citep{vol01}.  About 10\% of the stripped
material would then be in atomic hydrogen, the rest in molecular gas.

There are only a few observational constraints on the fate of stripped gas and it is
not clear what is more likely to happen under which circumstances: does the
gas end up being very hot or very cold? Only one clear case is known of a
trail of stripped gas in a cluster. In the cluster A1367,
\citet{gav01} have found a 75-kpc long trail of H$\alpha$ emission emanating
from two cluster galaxies, likely to be the result of a combination of tidal
interaction and ram pressure. It is not known what ionises this trail and
whether it has a neutral counterpart. Its extent suggests that the evaporation
timescale is at least $10^8$ yr.  Perhaps another clue is that some small,
isolated intra-cluster \HII\ regions exist in the Virgo cluster
\citep{ger02,cor04}. If these form from stripped gas, this would  
indicate that the cold scenario, at least sometimes, occurs.

   \begin{figure} \centering \includegraphics[width=8.5cm]{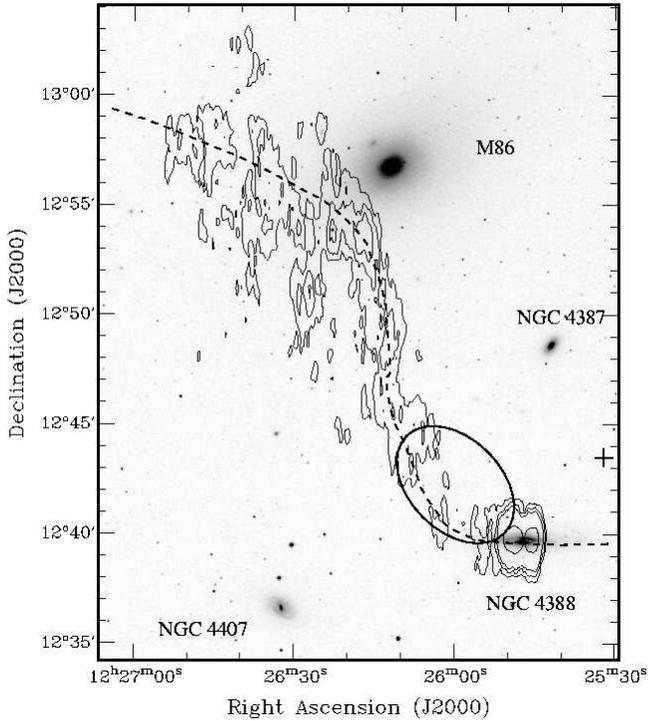}
   \caption{Integrated \HI\ emission of the plume. Contour levels are 1.0, 5.0,
   10.0 and 50.0 in units of $10^{19}$ \cmtwee. The locus along which the
   position-velocity plot of Fig.\ 2 is taken is indicated. The ellipse
   indicates where the ionised gas is detected  \citep{yos02},
   while the plus sign gives the location of the \HII\ region found by
   \citet{ger02}} \end{figure}

A prime candidate for further study is the Virgo spiral NGC 4388. This galaxy
is moving with a velocity of (at least) 1500 \kms\ through the Virgo cluster
and is one of its most \HI\ deficient galaxies (having
lost 85\% of its \HI; \citet{cay90}).  Its \HI\ disk is truncated to
well within the optical disk \citep{cay90}. Moreover, the galaxy has a large
plume of ionised gas, visible in H$\alpha$, extending 35 kpc off the galactic
plane to the  NE  \citep{yos02}. X-ray observations show the existence
of soft X-ray emission out to 16 kpc (and possibly 30 kpc),  in
a similar  location as the optically visible ionised gas
\citep{iwa03}.  Finally, by  comparing single dish and VLA
\HI\ observations, \citet{vol03} find evidence for neutral
gas out to at least 20 kpc NE of NGC 4388 with a \HI\ mass of $6\times 10^7\
M_\odot$.

Taken together, these observations indicate that the ISM of NGC 4388 is
strongly affected by the passage through the  cluster and that there is
large reservoir of gas out to at least 30 kpc NE of NGC 4388. No agreement
exists about what is responsible for the gas found NE of NGC 4388. Ram
pressure has been suggested \citep{vol03}, but alternative explanations have
also been proposed such as a nuclear outflow related to the Seyfert AGN in NGC
4388 \citep{vei99} or accretion of a small companion \citep{yos02}.  The
optical- and X-ray data suggest that the plume closest to the galaxy is
photo-ionised by the AGN residing in NGC 4388, but it is not clear what
ionises the gas further out.

Here we present new \HI\ observations of the region NE of NGC 4388. These
observations show that the gas NE of NGC 4388 extends much further than
previously thought. The data provide strong evidence that ram pressure pushed
this gas from NGC 4388.  Most likely it is not the ICM of the Virgo cluster
itself, but the hot halo around M86 that is stripping NGC 4388. Moreover,
stripped neutral gas is able to survive for at least $10^8$ yr. Locally, the
column density is possibly high enough for star formation to occur.

\section{The detection of a large \HI\ plume  NE of NGC 4388}

A region NE from NGC 4388 (pointing centre $\alpha$: $ 12^{\rm h}\,26^{\rm
m}\,00^{\rm s}$ $\delta$: $ 12^\circ\, 45^\prime\, 00^{\prime\prime}$, J2000)
was observed  for 12 hr with the Westerbork Synthesis Radio Telescope (WSRT) on
26 February  2005. A 20-MHz bandwidth was used, centred on a heliocentric
velocity of 1500 \kms, with 1024 channels and 2 independent polarisations. The
standard WSRT 72-m configuration was used. The data were calibrated following
the standard WSRT recipes using the MIRIAD software.  The datacube was made
with a channel width of 16.5 \kms. Hanning smoothing was applied to the final
datacube, giving a velocity resolution of 33.0 \kms. Robust weighting was used
with robustness = 0.5. Due to the low declination of the field and the E-W
layout of the WSRT the beam is very elongated : $18.0^{\prime\prime}\times
95.1^{\prime\prime}$ (P.A.\ $0^\circ$). The noise in the datacube is 0.27 mJy
beam$^{-1}$. This gives a 3-$\sigma$ mass sensitivity of $1.6\times 10^6\
M_\odot$ per resolution element (assuming a distance of 16 Mpc) and a
3-$\sigma$ column-density sensitivity of $1.7\times 10^{19}$ cm$^{-2}$. Cubes
with lower spatial resolution were also made, but these did not reveal any new
features. Hence, only the full-resolution data are discussed.

   \begin{figure*}
   \centering
   \includegraphics[angle=270,width=16cm]{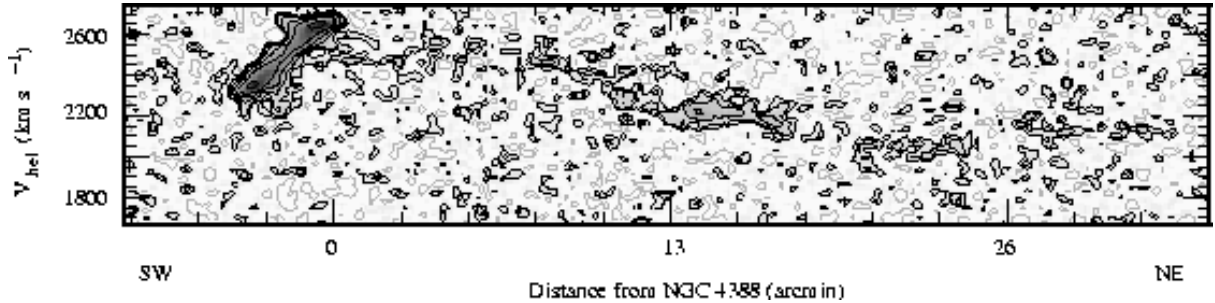}
   \caption{Position-velocity plot of the plume taken along the locus
shown in Fig.\ 1. Contour levels are --0.35, +0.35, 0.7, 1.4, 2.8, ... mJy beam$^{-1}$}
    \end{figure*}

\subsection{Properties of the plume}

In Fig.\ 1 the integrated \HI\ emission of the observed region is given,
superposed on an optical image taken from the Digitized Sky Survey.  The
integrated \HI\ image was made by selecting by hand those regions in the
datacube containing \HI\ emission.  This figure shows that we detect a large,
linear plume of \HI, starting from the E side of NGC 4388 and going in a NE
direction.  The velocity range of the plume is roughly 2000-2550 \kms\ which
is close to the systemic velocity of NGC 4388 (2524 \kms).  In Fig.\ 2 a
position-velocity plot is given taken along the ridge of maximum flux density
of the plume (indicated in Fig.\ 1).  The fact that the plume connects to NGC
4388 not only on the sky but also in velocity demonstrates that the plume is
physically related to NGC 4388. 

Given the pointings used in the single-dish \HI\ observations of
\citet{vol03}, it is clear that they detected the southern part of the
plume. Interestingly, \citet{dav04} have detected the \HI\ plume in their
survey of the Virgo cluster (their VIRGOHI4), however they interpreted
this emission as belonging to a galaxy behind M86 and hidden from our view.
Quite importantly, the \HI\ plume appears to be the neutral extension of
the plume of ionised gas detected by \citet{yos02} and \citet{iwa03}. 
The ionised and the neutral filaments clearly form one structure.

The extent of the plume is roughly 110 by 25 kpc and its \HI\ mass is
$3.4\times 10^8\ M_\odot$. This is about the same amount of neutral hydrogen
as we detect in NGC 4388 ($3.6\times 10^8\ M_\odot$). The peak of the
\HI\ emission is not located near NGC 4388 but instead is close to M86, about
10 kpc SSE of this galaxy. At this location, the observed column densities
reach values just above $10^{20}$ cm$^{-2}$. Given the large beam of our
observations, the column density will locally  exceed this value by a
significant factor. Near M86 the plume is also widest with a width of about
25 kpc.  Going from M86 towards NGC 4388, the column densities slowly
decrease and near NGC 4388 the plume is barely detected at levels of $10^{19}$
cm$^{-2}$. The distance from the maximum to NGC 4388 is about 70 kpc. The
plume extends for  another 40 kpc to the NE from the location of the
maximum. It is possible that the plume extends even further in this
direction. The tip of the plume is about $19^\prime$  away from the
pointing centre of the observations so  that primary beam effects
are  significant there (the FWHM of the WSRT primary beam is $35^\prime$).

The position-velocity plot shows that the velocity of the plume slowly
decreases from 2550 \kms, i.e.\ the velocity of the eastern side of NGC 4388,
to about 2000 \kms at the end of the plume. The velocity width is about 100
\kms, except in the region near the maximum where the velocity width is about
200 \kms.

\section{Discussion}

\subsection{Origin}

The morphology and kinematics of the plume clearly suggest that it is gas
pushed or pulled from NGC 4388, either by tidal interaction and/or ram
pressure stripping.  The relative velocities between galaxies in the Virgo
cluster are high, causing  tidal interactions between galaxies to occur
frequently, but to be short in duration.  The cumulative effect of many tidal
interactions (harassment; \citet{moo98}) can be significant, but one may
expect that harassment would not produce a single, long linear filament such
as the \HI\ plume, but instead a more diffuse or chaotic structure.  

An alternative is that the interaction is with  a larger scale
mass such as with a cluster sub-clump. Such interactions more likely produce long,
linear filaments because the timescale is longer. A nice example is a long
stellar trail in the Centaurus cluster. This trail is likely the
result of a tidal interaction of NGC 4709 with substructure of the Centaurus
cluster \citep{cal00}. The plume near NGC 4388 bears a striking morphological
resemblance to this stellar trail in the Centaurus cluster.  However, for this
kind of interactions to produce large tails, the interaction has to be
prograde \citep{cal00}, i.e.\ the internal spin vector of NGC 4388 has to be
more or less parallel to that of the orbit of NGC 4388 relative to the
substructure it is interacting with. However, the sense of rotation of NGC
4388 is such that, given the position and the large positive velocity of NGC
4388, any large-scale tidal interaction is retrograde, whether the interaction
is with the mass centred on M87, or the M86 sub-clump.

The main argument, however, against a tidal origin is that a
deep optical image \citep{phi82} shows that a stellar counterpart is
completely absent.  In general, given its larger extent, the \HI\ disk of a
galaxy is more susceptible to tidal distortions than the stellar disk.
However, given that the \HI\ disk of NGC 4388 is affected to well within the
optical disk, one would expect at least some stellar  counterpart to
the plume. The deep optical image does show  very diffuse stellar
emission all around  M86 and NGC 4388, suggesting that tidal
interactions have occurred several times in the past. Moreover, the image also
shows a small stellar feature going SW from NGC 4388. This suggests that NGC
4388 may have experienced minor tidal effects recently. These will not have
caused the  plume, but they may shaken the ISM of NGC 4388 such that other
effects, such as ram pressure, have had a more dramatic impact.

The lack of a stellar counterpart to the \HI\ plume, and the \HI\ disk being
truncated to well inside the optical disk, argue in favour of a ram pressure
related origin.  Indeed, NGC 4388 is often discussed as a good example of a
galaxy undergoing ram-pressure stripping
\citep{vol01,vol03}.  A fairly detailed model for the case where NGC 4388 is
interacting with the halo of hot gas centred on M87 is given by
\citet{vol03}. The observed morphology and kinematics of the plume match those
predicted by this model quite well (compare e.g.\ their Fig.\ 5 with our
Fig.\ 2).  However, another possibility is that the \HI\ plume is due to
stripping by the halo of M86.  Projection effects make an unambiguous
association impossible, but the plume appears to be much closer to M86 (10 kpc
projected distance) than to M87 (350 kpc projected distance).  The density
distribution along the \HI\ plume is what one would expect from an interaction
with M86.  The vicinity of the maximum density near M86, the fact that the
plume seems broadest there and the large profile widths, may suggest the
ram-pressure effects were strongest near M86. The kinematics of the plume is
well explained by the model of \citet{vol03}, i.e.\ going from NGC 4388 along
the plume, the velocities become less positive. However, the main reason for
this is that in the model NGC 4388 has a positive velocity w.r.t.\ the ICM
assumed to do the stripping.  Given that M86 has a negative velocity w.r.t.\
NGC 4388, similar gas kinematics would result from an interaction with the
halo of M86.  Detailed simulations would be needed to distinguish between the
two possibilities.

According to \citet{fin04}, the halo of M86 can provide the same ram pressure
as the halo of M87, so a passage of NGC 4388 close to M86 will have similar
effects as in the model of \citet{vol03}. In the model of \citet{vol03} for
the interaction with M87, the maximum pressure used is 5000 cm$^{-3}$
(\kms)$^2$. The relative velocity of NGC 4388 with M86 is much higher than
with M87, so the implied ICM density is lower: $4\times 10^{-4}$ cm$^{-3}$
vs. $1.25\times10^{-3}$ cm$^{-3}$. The current X-ray data do not allow to
estimate to which radius of the M86 halo this corresponds.

\subsection{Fate of gas}

Different predictions are made in the literature for what happens to the gas
once it has been stripped. The large extent of the \HI\ indicates that it can
survive for a fairly long time. Although projection factors are unknown, a
rough estimate of the age of the plume can be made.  Assuming NGC 4388 is
moving in the plane of the sky with a velocity of 500 \kms, the length of the
plume implies an age of a few times $10^8$ yr. This may suggest that
evaporation by the ICM is slow, e.g.\ because the heat flow is saturated
and/or that a tangled magnetic field slows down the heat flow into the plume
\citep{vol01}.  The X-ray halo of M86 is much smaller in size than that of M87
so the ICM density drops faster with radius, while also the higher relative
velocity of NGC 4388 with M86 implies lower densities for the region
responsible for stripping NGC 4388.  This could mean that if the plume is due
to stripping by M86, longer evaporation times would then be expected.
Given the long evaporation timescale, dense clumps may form in the plume and
part of the stripped ISM may become molecular.  \citet{vol01} predict that in
such a scenario only 10\% of the total gas mass may be in the form of neutral
hydrogen. In this  case, the total gas mass of the plume is
$3.4\times10^9\ M_\odot$.

In and around normal spirals, if the \HI\ column density is above a few times
$10^{20}$ \cmtwee, star formation almost invariably occurs \citep{sch04}.  The
column densities in the plume likely exceed this threshold in some locations,
in particular near M86.  Hence, star formation could occur locally in the
plume, provided the processes that regulate star formation for a cloud in the
ICM are similar to those for gas clouds in spiral galaxies.  Indeed, at least
17 extragalactic \HII\ regions exist near M86 and NGC 4388
\citep{ger02,cor04}.  We do not detect \HI\ near the \HII\ region found close
to NGC 4388 by \citet{ger02} so there may not be a direct connection for this
particular \HII\ region.  The number of intra-cluster \HII\ regions is small,
so most of the mass of objects such as the plume will end up being part of the
hot ICM and will enrich it.

An interesting question is how common  neutral gas clouds, as the one we
observe near NGC 4388, are.  There are many Virgo spirals currently undergoing
ram-pressure stripping.  If the stripped gas remains observable for at least
$10^8$ yr, more plumes should exist.  Many Virgo spirals have been imaged in
\HI\ and no large plumes have been reported.  However, our
data  show that new,  deeper observations may reveal plumes in places
where it is thought they do not exist.  Other recent  observations
also start to show much more detail in the \HI\ distribution of stripped galaxies
\citep[e.g.][]{chu05}. Moreover, in a blind single-dish survey of the Virgo
cluster, \citet{dav04} found 3 \HI\ clouds (of which the plume of NGC 4388
turns out to be one) that do not have an optical counterpart.  So it appears
that a few more \HI\ plumes may exist in the Virgo cluster.  
\citet{min05} claim that one of the clouds found by \citet{dav04} could be a
``dark galaxy'', i.e.\ an object containing gas and dark matter, but no stars.
However, located about 200 kpc from this cloud is the galaxy NGC 4254, a
spiral that shows signs of both tidal interaction and ram-pressure stripping
\citep{pho93}.  The velocity of the gas cloud and NGC 4254 are very similar,
so it is quite possible that this cloud has a similar origin as the plume near
NGC 4388.

\begin{acknowledgements}
We would like to thank Bernd Vollmer for his valuable input.  The WSRT is
operated by the Netherlands Foundation for Research in Astronomy (ASTRON) with
the support from the Netherlands Foundation for Scientific Research (NWO).
This research has made use of the NASA Extragalactic Database (NED). The
Digitized Sky Survey was produced at the Space Telescope Science Institute
under US Government grant NAG W-2166

.

\end{acknowledgements}

\end{document}